\documentclass[aps,reprint,amsmath,longbibliography]{revtex4-1} 

\bibpunct{}{}{,}{s}{}{}

\usepackage{dcolumn} 
\usepackage{graphicx} 

 \newcommand{\figwide}{7.8cm}

\usepackage{array} 
\newcolumntype{K}[1]{>{\centering\arraybackslash}p{#1}} 
\usepackage{longtable}

\usepackage{color}					

\begin{document}

\title{Students' epistemological framing in quantum mechanics problem solving}

\date{\today}

\author{Bahar Modir}
\author{John D. Thompson}
\author{Eleanor C. Sayre} 
\email{esayre@ksu.edu}
\affiliation{Department of Physics, Kansas State University, Manhattan, Kansas 66506}

\begin{abstract} 

 Students' difficulties in quantum mechanics may be the result of unproductive framing and not a fundamental inability to solve the problems or misconceptions about physics content. We observed groups of students solving quantum mechanics problems in an upper-division physics course. Using the lens of epistemological framing, we investigated four frames in our observational data: algorithmic math, conceptual math, algorithmic physics, and conceptual physics. We discuss the characteristics of each frame as well as causes for transitions between different frames, arguing that productive problem solving may occur in any frame as long as students' transition appropriately between frames. Our work extends epistemological framing theory on how students frame discussions in upper-division physics courses.

\end{abstract}


\maketitle


\section{Introduction}

For students to be successful quantum mechanics problem solvers, it is insufficient to think about only the features of the physical system. They also need to coordinate different representations by thinking conceptually about the mathematical representations that satisfy the physical system, evaluate the algorithmic steps, and reflect upon their work. Unsurprisingly, students often have trouble unifying these ideas during problem solving. 

Researchers in student understanding of quantum mechanics have used ``difficulties" theory to understand student reasoning (e.g. \cite{Singh2015, Emigh2015, Passante2015}), which forms long lists of difficulties that span many topics in quantum mechanics. 
However, we posit that these disparate difficulties can be unified through the lens of epistemological framing\cite{Tannen1993}, and errors in transitions between frames\cite{Irving2013framing}.  Epistemological frames reveal students'\cite{Redish2004, Tannen1993} ways of thinking and expectations. They govern which ideas students link together and utilize to solve problems. Careful observation of student behaviors, gaze, and discourse can provide clues for determining students' epistemological frames. Productive problem solving requires both an appropriate frame \cite{Scherr2009} and appropriate transitions between frames\cite{Nguyen2016StuFraming}.  

Some students, despite having strong numerical tools or skills, still  ``get stuck'' in certain problem solving situations\cite{Tuminaro2004a}.  This happens particularly in upper-division courses such as quantum mechanics, where mathematics is critical to understanding the subject. Quantum mechanics is a great choice for this study, because students are trying to coordinate difficult, often counter-intuitive concepts and complicated, often novel mathematical formalism.    

Bing et al\cite{Bing2012} identified four epistemological frames to aid in understanding the role of math as a reasoning tool as opposed to a numerical tool. They analyzed students' thinking while the students translate physical ideas into informative mathematical forms, or compare a mathematical structure in two similar physics or math scenarios. Though Bing et al identified ``Physical Mapping'', and ``Math Consistency'' frames, their ``Calculation'' frame is biased toward the use of formal math, independent of physical sense making. However, they did not further differentiate between trivial math calculations and conceptual math reasoning.

On the other hand, Kuo et al \cite{Kuo2013Blending} differentiated between the use of equations as an input-output calculator template, instead of attending to the conceptual meaning embedded in the equation to create shortcuts. They referred to ``cognitive elements'' \cite{Sherin2001} to capture students' understanding of equations while they blend their reasoning with symbolic forms and create a shortcut to interpret the situation. They concluded that successful problem solvers are able to make a decision as to which tools they bring into play for an efficient understanding of the problem situation.

Earlier studies identified other possible avenues that students may follow to obtain a correct solution by using conceptual physics and algorithmic math as numerical tools.  These problem solving studies often focused on the differences between experts and novices\cite{Larkin1980, Heller1997, Chi1981a}.  A novice adopts an inverse strategy by simply attending to the goal of the problem, recalling and manipulating an equation that contains the unknown quantities. In contrast, an expert moves forward based on having a representation for the situation, and then choosing the relevant principles\cite{Larkin1980}. However, this study is limited because experts' expertise far exceeds the difficulty of the end-of-chapter problems, and so such a study can not show the heuristics of expert-like problem solvers.  

Heller et al \cite{Heller1997} designed context-rich problems to challenge introductory students beyond end-of-chapter exercises. This method requires students to make sense of the physical system, and justify what strategies to adopt as experts do. Their work assesses how students initially translate the problem statement into a visual representation in order to help them to adopt a proper strategy for determining the implicit unknown physical quantities. The strategy would allow students to translate their physical representation into a mathematical representation in order to do the algorithmic steps, and find the unknown quantity to make sense of their solution.  While this problem solving strategy was initially quite prescriptive in the nature and order of the problem solving steps, later research has permitted a less-linear structure to problem solving. 

Building on this work,  Caballero et al \cite{Caballero2013ACER} worked to explain the common difficulties of upper-division students in problem solving, focusing on four steps in mathematical tool use: activation, construction, execution, and reflection.  These steps could be completed in any order, and solutions may vary among students and problems. One student could stay mostly in the execution phase to process the algorithmic steps. Another student might evaluate the solution by staying in the construction phase and skipping the execution elements in favor of conceptual steps, or one could bring into play both components of execution and construction.  Students' use of certain steps in this theory does not necessarily imply difficulties with the missing component of their problem solving process. This could become important when the problem statement of the question nudges students toward the use of one of these four components more than the others. 

Broadly speaking, these three research traditions -- research into student difficulties in quantum mechanics, research into epistemological framing, and research into student problem solving -- suggest several approaches for understanding how students understand quantum mechanics problems. One approach may consider the initial physical understanding of the problem as more critical, with less emphasis on the mathematical manipulations, whereas other approaches may consider equally a close relationship between math and physics, or focus on the conceptual meaning of math in reasoning. Our present study integrates these three approaches to capture the various facets of students' epistemological framing during problem solving at the upper division.  Our theoretical framework\footnote{It is a particular linguistic difficulty of research on students' epistemological frames that the theory used to describe them is a ``framework''.  In this paper we follow the convention that a ``framework'' is something that researchers use, while a frame is something that humans (in our case, students) use.  A framework is an appropriate technical term for a set of connected theoretical statements (e.g. "Resources Framework") Students -- humans -- frame ideas, have epistemological frames, and participate in framing activities.  There are some subtle differences between these three forms of "frame", but all of them are related to the idea "how you know what's going on". }
takes up the idea of epistemic frames to explain student problem solving without prescribed steps. Our model suggests that difficulties are an interaction effect between question asked and students ideas, which implies there may be an underlying structure to identified difficulties in quantum mechanics.

In this paper we develop a theoretical framework which models students' framing in math and physics, expanded through the algorithmic and conceptual space of students' problem solving. We investigate four frames: algorithmic math, conceptual math, algorithmic physics, and conceptual physics, looking for moments where students' problem solving is impeded because they are in an unproductive frame.  We applied this theoretical framework to observational data from quantum mechanics classes in which students solve typical quantum problems in pairs and small groups. Our purpose in this paper is to illustrate our theory, not to exhaustively show the prevalence of specific frames or to catalog the methods by which students may transition between them.

\section{Context}

We video recorded the class meetings of one semester of a senior-level quantum mechanics class. The class is taught using Griffith's Introduction to Quantum Mechanics\cite{Griffiths2005QM} using a wavefunctions-first topic order.  It meets for four 50-minute sessions each week.  During class, lecture is interspersed with small group problem solving.  Groups of 2-3 students solve problems collaboratively on shared table-based whiteboards. Most problem-solving sessions last 2-5 minutes, though they can be as long as 15 minutes for more difficult problems. Students are remarkably collaborative, usually working together for the entire duration of each problem-solving session.  In our data set, we see about one problem-solving session per class, though this decreases in frequency near the end of the semester.  

Generally, these problem-solving sessions begin when the professor halts the lecture to ask the students to attempt to solve a problem related to their current topic, or to introduce a new topic. Occasionally, they also arise when students initiate a class discussion and the professor decides to assign a problem to gauge their understanding. 

The groups in this class are somewhat fluid, and students may form different groups on different days.  Students occasionally recruit others from nearby groups to help them solve problems.  The instructor does not explicitly tell students where to sit or with whom to work (other than ``people near you'').  Generally speaking, students work in pairs or threes; occasionally fours.  

\section{Methodology for video data}

In learning environments such as group problem solving in upper-division contexts, one way to interpret the high level of interactions within group members is to carefully analyze the discourse and gestures of each member of the group. Ethnography provides an opportunity to understand the detail of students' discourse, behaviors, as well as capture useful information while they are investigating a phenomena\cite{Pirie1997}. One of the methods for data collection in ethnography studies in through video recording of activities. This becomes more important by providing multiple researchers an opportunity to view and analyze the videos\cite{Derry2010}. Previous researchers in education have used ethnography to study the culture of classroom activities \cite{Brown2004} or in more engaging learning environments, such as advanced physics laboratory \cite{Irving2014AdLab}.  Our goal was to develop a theoretical lens to enable us to explain problem solving within various topics in quantum mechanics. 

We divide class into episodes of problem solving and episodes of lecture, discarding episodes of lecture because they don't help us understand student reasoning.  The problem solving episodes have distinct boundaries: they start with the professor explicitly asking students to begin working on their table-based whiteboards and end when the professor either asks for answers or begins explaining the answer.  

In our preliminary analysis of the students' group problem-solving activities, we observed that some aspects of the data represent a conceptual approach and other aspects represent an algorithmic approach. We also noticed students' use of conceptual physics and algorithmic math. This distinction is consistent with the ACER\cite{Caballero2013ACER} and framing\cite{Bing2012} literature on problem solving from upper-division physics classes, showing how students' understanding of physical systems maps to algorithmic representations. However, neither theoretical framework adequately captured the richness of our data, prompting us to take further steps to interpret our data set.  From the tradition of progressive refinement of hypotheses\cite{Engle2007}, we set out to refine our observations through close interrogation of the video data.

We started with selecting episodes for close analysis based on their duration (longer is better), conceptual richness (more complex is better), and technical quality (more visible and audible are better).  We reflected on these episodes, seeking to answer ``what's going on?'' for each of them. Through repeated watching and examining the details of the selected episodes, we sought to capture changes in students' discussion or behavior that might indicate a shift in the students' problem solving processes.  We began to focus on instances where students ``got stuck'' in their problem solving processes. This momentary impasse prompted them to try a different kind of reasoning until suddenly they were able to get ``un-stuck''. We examined the interactions immediately preceding and following the unsticking moments to look for regularities in unsticking behavior.

We developed a preliminary theoretical framework\cite{Thompson2016ICLS} which mapped student behavior onto three discrete frames: conceptual physics, conceptual math, and algorithmic math.  The two math frames -- concordant with research in mathematics education on concepts and processes\cite{Sfard1991,Dubinsky2001} -- suggested that we expand our ideas to look for the ``missing'' physics frame: algorithmic physics.  

Concurrently, we grew troubled with the idea of discrete frames.  Sometimes, students seemed exceptionally ``mathy'', operating without regard to any sense of physical meaning.  It is possible, however, to blend conceptual ideas from both math and physics domains, or to move fluidly and rapidly between conceptual and algorithmic thinking.  We reframed our ideas into two orthogonal axes: conceptual to algorithmic and math to physics, defining a coordinate plane in which students' problem solving roams.  In pursuit of evidence to refine this two coordinate-axis framework, we delved again into our observational data, seeking examples of all four quadrants and transitions among them.  

After several more iterative cycles of analysis and refinement, we reached a stable point where new episodes did not change the theoretical framework or our application of it. Operating with the newly-stable framework, two independent raters came to consensus on every episode; two additional raters checked a selection of episodes with agreement of $>90\%$. We selected episodes for analysis based on frequency of students' discussion regarding concepts and processes, as well as displays and features of potential frame transitions. We categorized episodes with conceptually rich discussions and frame negotiations as strong examples, and established inter-rater reliability about the content of the episodes and regarding which episodes strongly or weakly evidenced frame transitions. 

We also identified very weak examples when it was hard to find evidence of students' framing from the group discussion. This could happen due to noisy or garbled audio, or when students were writing on part of the whiteboard that was not in the view of the camera, or in general the raters did not have enough information to determine students' framing.

Once we identified students' frames, we looked for transitions in those frames to help us to interpret the dynamic of students' problem solving behaviors or identify the impasse students reach when they fail to notice certain factors that could have triggered a transition to a more appropriate state.

We acknowledge the existence of other frames that could describe students' behavior while they are engaged with other kinds of activities in a classroom e.g. turning in home work to the instructor, taking break within solving several parts of a long problem, or discussing upcoming social events\cite{Nguyen2016Dynamics}. While these other frames can be important for problem solving more broadly\cite{Irving2013framing}, in this study our focus is on investigating students' topical discussions during problem solving sessions which last about 2-5 minutes.  

Epistemological frames are context dependent\cite{Redish2004}. For example, by walking into a restaurant relevant resources consistent with behaving in the situation are activated to read a menu, order food, pay the tip, etc. However, in the setting of the restaurant we don't access our resources for behaving in a library. 
Students' perceptions of the problem context affect their framing of what subset of their knowledge to activate. In an interactive class environments such as group problem solving the instructor's framing can also affect the students framing of the situation\cite{Scherr2009, Irving2013framing, Nguyen2016Dynamics}. Even within group problem solving, other students' framing can affect an individual's framing as well. The instructor can nudge students to frame the problem more conceptually by asking about the physics of the situation, or more algorithmically by asking about formulae\cite{Nguyen2016Dynamics}.

\section{Theoretical framework}
 
\begin{figure}
 \includegraphics[width=\figwide]{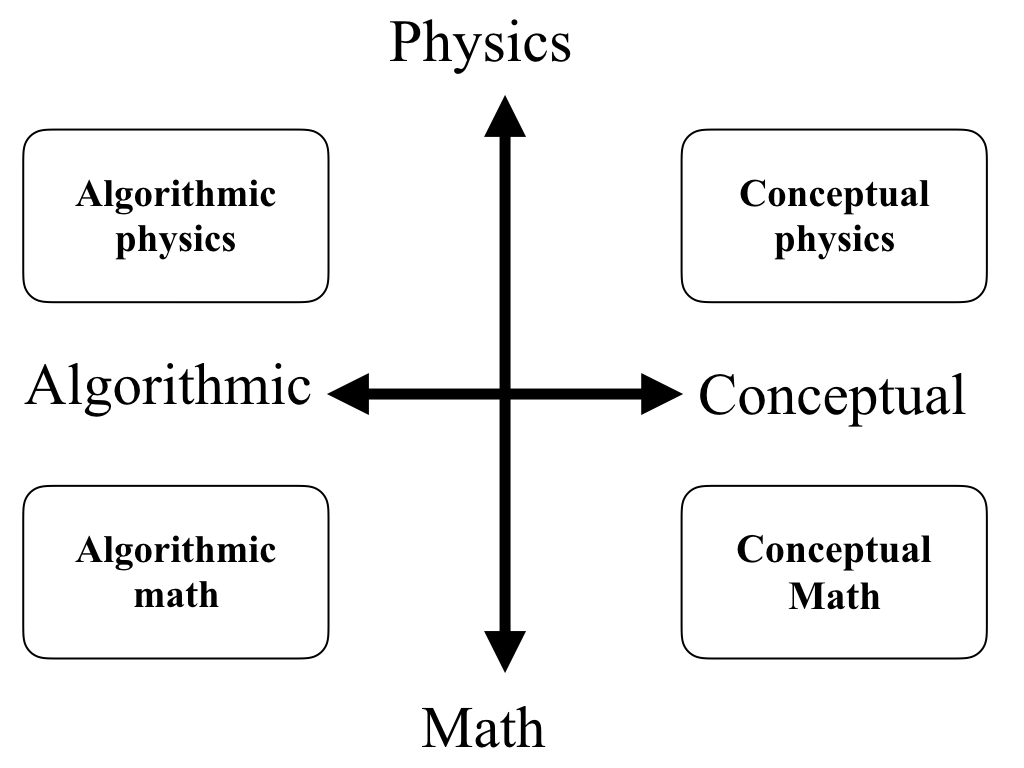}
 \caption{Math-physics-algorithmic-conceptual theoretical framework. The horizontal axis indicates algorithmic and conceptual directions. The vertical axis represents the math versus physics directions. Each quadrant is labeled.\label{fig:framework}} 
 \end{figure}

 Our theoretical framework consists of two axes: an algorithmic versus conceptual axis, and a math versus physics axis (Figure \ref{fig:framework}). 

The two axes divide different aspects of students' problem solving into four regions: algorithmic math, conceptual math, algorithmic physics, and conceptual physics. It is important to note that none of these frames are inherently ``good'', ``bad'', or even universally useful.  At different times, different frames may be productively used to solve problems in physics, and often more complicated problems require multiple transitions between frames. 



\subsection{Algorithmic and conceptual math}

In algorithmic frames, students are focused on following a known series of steps to solve a problem. Since the result of each step is used to process the following step, students are focused on their task to prevent errors. They tend to be in a writing mode, and have less discussion as compared to conceptual frames. In their discussions, they tend to focus on error-checking (``what did you get for part c?'' or minutiae of their steps (``you are missing a minus sign''). 

When the problem statement requires explicit algorithmic calculations to find an answer, students enter an algorithmic math frame to spend a considerable amount of time setting up a series of algebra-based steps to evaluate integrals, take derivatives and simplify their solutions by dividing or multiplying a term on both sides of their solution.  

Algorithmic math can be a quick and powerful problem-solving mindset as they may take several fast steps over a long period of time. However, without other quadrants it is quite difficult to check whether or not the solution makes sense. This frame leaves students with a narrow\cite{Irving2013framing, Engle2012Expansive} discussion mostly to check the signs, or to alert each other of the missing symbols, while they are focused in their numerical calculations. For these features, we consider the ``just math'' frame\cite{Wolf2014JustMath} is an example of students' prolonged use of the algorithmic math frame.  Narrow framing which focuses only on the task at hand\cite{Engle2012Expansive, Irving2013framing} is not exclusive to algorithmic frames, but it is common within them.  

In contrast, students in a conceptual math frame use a conceptual approach to understand the mathematics. They reason based on general properties of a class of information in math. This could help them to apply practical ideas about the behavior of the mathematical functions, and determine the result of an operation without actually computing it. For example, by knowing that sine functions are independent of each other, and discussing the orthogonality properties of the sine functions, one can shortcut the integration of product of two sine functions of different periods, preventing the use of many trigonometric identities, and simply ``see'' that the integral equals zero.
 Creating a ``shortcut'' solution\cite{Kuo2013Blending} to the problem reduces the procedures and lessens the writing.  Concurrently, discussing mathematical problems conceptually gives students more opportunities for sense-making discussions with other members of the group\cite{Scherr2009, Irving2013framing}.  This kind of thinking is generally more expansive\cite{Engle2012}, as students connect general cases of mathematics to the specifics in this problem or bring in connections to other problems. 
Attention to conceptual mathematics is a large part of numeracy, and as such is an important part of learning mathematics\cite{Dubinsky2001} and physics\cite{Kuo2013Blending}, especially at the upper-division\cite{Sayre2008,Kustusch2013}.


\subsection{Algorithmic and conceptual physics}
Just as we find algorithmic and conceptual frames in math, we find them in physics as well. Students in a conceptual physics frame try to think in terms of the features of the physical system and might coordinate between different representations such as graphical, geometric or gestural to visualize the physical system. They coordinate different physical laws and concepts to explain the situation. We provide examples from the context of Electromagnetic fields course as motivating examples to show the broader phenomena. In the next section, we will provide several quantum mechanics examples from our own data.  

For example, students might argue that the total charge on a spherical shell whose surface charge density $\sigma$ is proportional to $\sin(2\theta)$ (where $\theta$ is the azimuthal angle) is equal to zero because the northern hemisphere is positive while the southern is negative, and those two halves must be equal and opposite.  In this case, students use conceptual reasoning to map charges to a sphere, employing balancing resources to come to the conclusion that the net charge equals zero. One could move to algorithmic math frame to write the integral of the charge density over the surface area: 
\begin{align}
\int{\sigma}{dA}{\propto}{\int{\sin(2\theta)r^2\sin(\theta){d\theta}{d\phi}}}
\end{align}  
It's possible, of course, to solve this problem algorithmically (using trigonometric substitutions) or conceptually without reference to physical systems (via the orthogonality of sine functions).  In either the conceptual math or conceptual physics frames, discussing the problem plan in the conceptual physics frame can make later algorithmic calculations easier. 

Thinking conceptually about the underlying physics of the situation encourages students to create connections to real word situations and other classes of problems as well. For example, to estimate the far distant electric field of a uniformly charged disk, one method is to expand the solution by mostly engaging in algebra to get the answer. Or one can visualize that far from the disk a continuous charge looks similar to a point charge, and by knowing the electric field of a point charge, the leading terms in the solution can be guessed. In each case, students are engaged in an activity to find an answer, but the nature of the activities are different. The latter case is more expansive, as students are open to make connection between the current situation and another class of problem.

In contrast, students in the algorithmic physics frame tend to recall equations, facts, and properties of physical quantities without conceptual justifications. They use math as a tool to adjust equations via a series of algebra-based steps to relate physics quantities to each other, or to check the correctness of the physical quantities in the problem. For example, by doing dimensional analysis students can check the correctness of their answer. Just as with algorithmic math, students tend to frame their work narrowly in algorithmic physics and focus on following procedures to find answers.  Someone who applies normalization conditions for wavefunctions by rote, for example, is operating algorithmically.  

It's important to note that framing problems algorithmically can be fast.  An expert doesn't need to engage in extensive conceptual thinking about the steps of a trivial problem; she can just solve it.

\subsection{Continua vs. categories}
A careful reader might be concerned because we started this section by claiming that there are two axes, implying a continuous distribution of possible framings, yet continued by identifying four frames which appear to be discrete. We chose the axes for theory-driven reasons: it's possible that students' framing exists on a continuum between very mathy and very physicsy, or very algorithmic to very conceptual, and discrete frames cannot capture this sense.  We kept it for practical reasons: on occasion, students appear to move fluidly and rapidly among frames, and there's not enough evidence to assign them a single, quasi-stable frame before they move to the next.  We're interested in quasi-stable frames because we want to study how students transition between frames, and it is practically very difficult to find transitions between frames without first identifying (quasi-)stable frames. We use the words mathy and physicsy to denote students' framings which are more in the math direction or which are more in the physics direction respectively. 


By using axes, we hope to capture a sense of directionality from more mathy to more physicsy and more conceptual to more algorithmic. We do not imply that these axes constitute a formal metric or scale.  
While some prior work in student framing of problem solving in physics has used discrete frames (e.g. \cite{Scherr2009,Bing2012}), other work has used continua in the same way (e.g. \cite{Irving2013framing}, building on \citep{Engle2006}).

 \begin{figure}
 \includegraphics[width=\figwide]{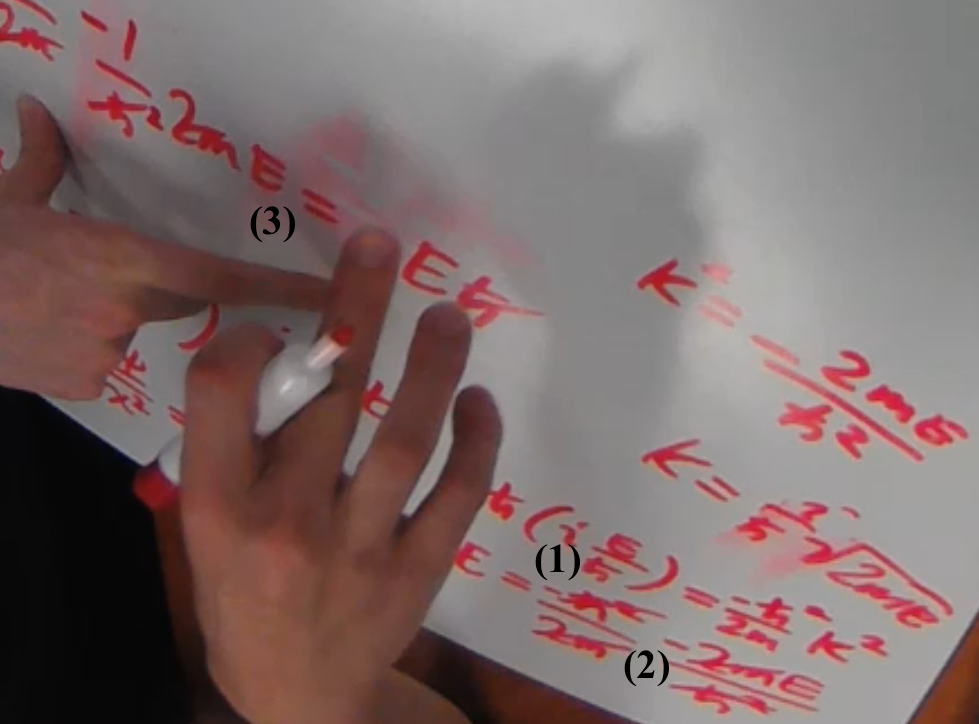}
 \caption{Group problem solving in algorithmic math frame\label{fig:snap}} 
 \end{figure}
 
  \begin{figure}
 \includegraphics[width=\figwide]{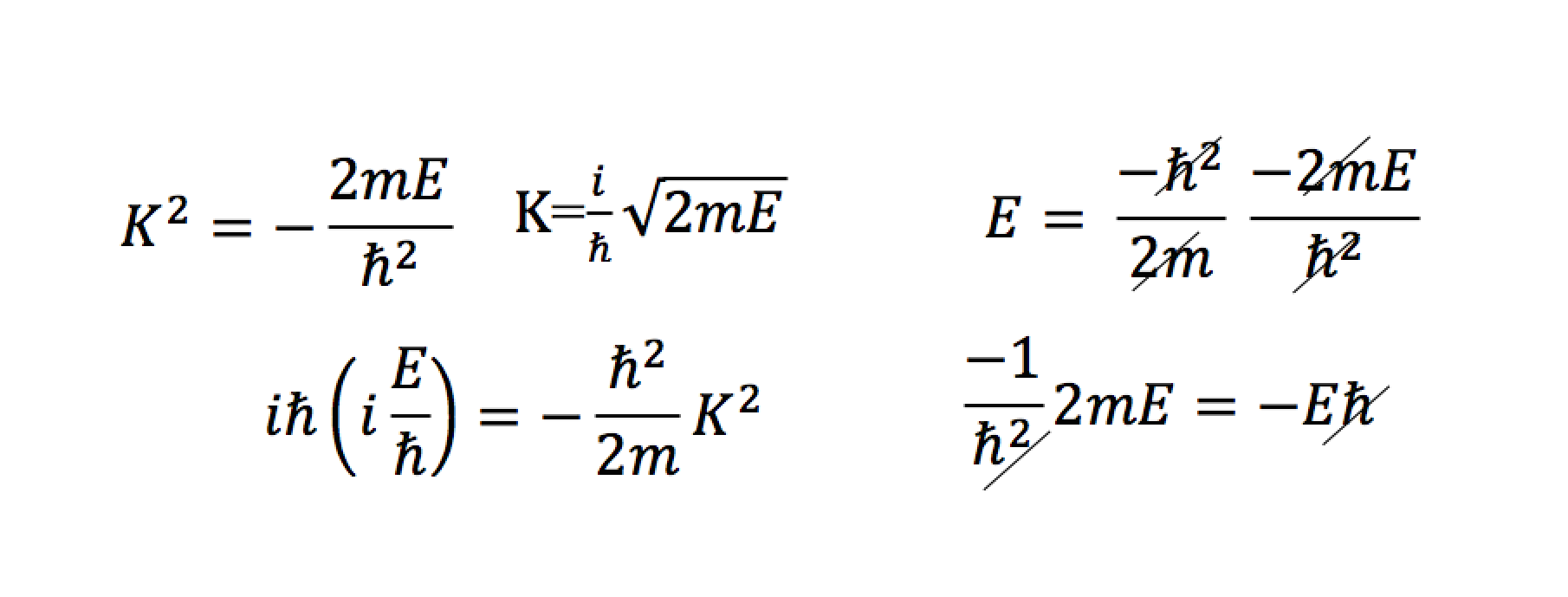}
 \caption{Diagram of students' solution in algorithmic math frame\label{fig:diag}} 
 \end{figure}

\section{Illustrative episodes}
In this section, we present four brief episodes which illustrate the four quadrants in our framework.  Before the examples, we divert into a brief review of the quantum mechanics of free particles and a typographic note on how we present transcript.

\subsection{Physics of free particles}


Most of the examples in this section are chosen from the same physics context of the free particle system, so we review the physics of this system briefly for the reader.  In quantum mechanics, the free particle is characterized by a zero potential energy,  thus the Hamiltonian has just one term, the kinetic energy. This problem is a good candidate for understanding basic properties of the wave function and the Schr{\"o}dinger equation. Since the Hamiltonian is in form of $p^2/2m$, the eigenfunction solution can be considered as a plane wave, which can be expanded in terms of sinusoidal wave functions. However, for one particle the wave function with a determined momentum is not normalized over all the space. Thus a linear combination of all solutions is considered as a normalizable wave function.  
This is physically interpreted as a traveling wave pocket.  

Because the Hamiltonian has only the kinetic energy term, the time independent Schr{\"o}dinger (TISE) equation results in a homogeneous second-order differential equation. In order to write the eigenvalue equation in terms of a differential equation students might need to recall some relations from algorithmic physics. Solving and finding the eigenfunctions of the equation leaves room for either algorithmic math calculations, or conceptual math discussions. 
 

\subsection{Typographic note}

Before we present data, here is a brief typographic note.  In these interactions, students very frequently speak the names of mathematical symbols.  We could have typeset their words as if they were equations or as if they were the names of isolated symbols.  Equations are more compact -- importing algebra to Europe caused a scientific revolution -- but they lose some of the nuance of students' speech.  Isolated symbol names, on the other hand, tend to be difficult to follow in text in a way that they are not difficult to follow in speech, especially as oftentimes students write as they speak.  We have chosen a middle path, seeking to maximize clarity for the reader.

Additionally, we typeset a comma for brief pauses, a period for longer ones, and ellipses (\dots) for the longest ones.  Stage directions are denoted by parentheses.  Should we omit or alter some students' speech for clarity, changed words are denoted by square brackets and omitted ones by ellipses in square brackets ([\dots]).

\subsection{Episode: Algorithmic math\label{sec:AM}}
In this example a group of three students are solving the Schr{\"o}dinger equation to find the wave function of a free particle. They treat the space part and the time part separately. They start with the TISE for the space part. Guess a solution in the form of $e^{kx}$, and substitute it into the TISE to find the constant $k$.
Adam and Emma work together quickly to solve the problem, while their third groupmate (Eric) stays silent. (All student names are pseudonyms.)

Figure \ref{fig:snap} shows a snapshot of their whiteboard during group problem solving, taken while Emma is pointing to the both sides of their written equation to review the taken algorithmic steps in search for the missing sign. The numbers indicated on the figure show the order of the students' actions and narrations and reference the numbers in the transcript below. Figure \ref{fig:diag} shows a transcript of their writing on the whiteboard. 

In this problem, Adam takes the derivative of the time solution of the Schr{\"o}dinger equation, and replaces it into the equation using the whiteboard in front of both students. He continues to replace the factors that the group has manipulated earlier in their solution and setting both sides of the equation equal to each other to verify if their solution satisfies the Schr{\"o}dinger equation. 

\begin{description}
\item[Adam] Minus $E$ equal
\item[Emma] Minus $\hbar$ squares over two. (1)
\item[Adam] Minus $2mE$ over $\hbar$ \dots squared \dots Boom \dots Boom \dots Boom (while canceling the same quantities from two sides of the equation) (2)
\item[Emma] Cancel, cancel, cancel, and we are off by a negative (2)
\item[Adam] Yeah [unintelligible] sign
\item[Emma] With a negative up here, because these two are negative.
\item[Adam] Yeah that is true
\item[Emma] So something happened here (pointing to the two sides of their equation) (3)
\item[Adam] Or we lost a sign (3)
\end{description}

Emma and Adam use short sentences and talk quickly to be able to proceed to the next step of their algorithmic evaluation. They speak primarily of mathematical terms and operations, and do not talk explicitly or extensively about the physical quantities these symbols represent.  At the end, they come up with an extra negative sign in one side of their solution. After reviewing their solution, Adam removes a negative sign in the earlier line of his solution, which he thinks is extra, but he does not further discuss the reason behind his decision. 

At this point of their problem solving session, neither Adam nor Emma try further to make a transition to another frame to resolve thier error. In a low voice, Adam points to the power of the time phase factor exponential and says ``oh wait this [$e^{i\frac{E}{\hbar}t}$] gonna be a negative''. Emma says ``but I think we can start normalizing\dots'', then she starts to normalize their wave function.

In this brief moment of problem solving session the group is in an algorithmic math frame. Students are operating in algorithmic math frame because they are talking about manipulating mathematical symbols rapidly and alegebraically, running through a series of brief steps.  
Prior to this episode, Adam makes a transition from algorithmic math to algorithmic physics.  In so doing, he is able to resolve the cause of their group error.  After, he returns to algorithmic math to continue the solution. (discussed in section \ref{sec:Adam})

The group does not spend further time to find the ``dropped negative" since students are toward the end of their problem solving session and are asked to normalize the wave function, which is discussed in the next episode.

\subsection{Episode: Conceptual math}

In this example, the same group is working on normalization of the free particle wave function by considering the general  solution $\Psi$ as the sum of two functions of $Ae^{ikx}$ and $Be^{-ikx}$. They initially set the algorithmic steps, take the modulus square of the wave function ($\mid\Psi^*\Psi\mid$) and insert the limits of the integral.  

\begin{widetext}
\begin{eqnarray}
\int_{-\infty}^{\infty} \mid\Psi^*\Psi\mid dx =& \int_{-\infty}^{\infty} \mid A\mid^2 + \mid B \mid^2 + AB^*e^{-2ikx}  +  A^*Be^{2ikx} dx \label{eqn:1}\\
=& x( \mid A\mid^2 + \mid B \mid^2) \Biggr|_{-\infty}^{\infty} 
+ \int_{-\infty}^{\infty} AB^*e^{-2ikx} dx +  \int_{-\infty}^{\infty} A^*Be^{2ikx}dx \\
=& AB^*\frac{1}{-2ikx}e^{-2ikx}\Biggr|_{-\infty}^{\infty} 
+ A^*B\frac{1}{2ikx}e^{-2ikx}\Biggr|_{-\infty}^{\infty} \label{eqn:2}
\end{eqnarray}
\end{widetext}

There are a few errors in the students' solution which will not affect their conceptual discussion later in the episode. Emma leaves just one differential elements of length $(dx)$ for the whole expression right after the last term in the integrand (Equation \ref{eqn:1}). However, Emma is mindful of her incorrect notation, as she will take the integral of each term  of the integrand with respect to $x$ in the following lines. The other error takes place after taking the integral of the $e^{2ikx}$: Emma leaves an extra negative sign in the power of the exponential, and none of the group members notice the extra sign for the second exponential integral (Equation \ref{eqn:2}). Eric notes that the result of the constant integrals are infinite; Emma will not further discuss and skip those terms in the last line of the solution.

They end up with the variable $x$ in both the denominator and the power of the exponential function in the numerator of the fraction (Equation \ref{eqn:2}). We acknowledge that the frame of students is mixed ahead of time while they are setting up their integrals algoritmically. But after this setup they switch to a purely conceptual reasoning and start their discussion again. 

Without evaluating the integral numerically, they realize that the answer of the integrals might be infinite. They start in a conceptual conversation in the math context by arguing based on words and properties of the wave functions rather than working out equations to justify their answer as being finite.

\begin{description}
\item[Emma] We don't have to worry, this [the exponential term in the numerator] is gonna blow up faster than this [the denominator], right! 
\item[Eric] They both blow up
\item[Emma] Yea, But one blows up faster and that matters
\item[Adam] Definitely the exponential (points to numerator)
\item[Emma] Yeah\dots
\end{description}

Emma has a discussion about which term ``blows up'' faster than the other. Emma gives more evidence of her conceptual understanding of the behavior of the two functions in the second exponential integral, when Eric says ``but both terms blow up''. She then compares the decay rate of the functions as an important factor that ``matters''.  Emma conveys her generalized expectation\cite{Sayre2015BESM} of the situation by saying  ``We don't have to worry''. 

Although Emma only uses the  term ``blow up'' briefly, there is a conceptual meaning embedded in this phrase which shows her understanding of the situation by comparing the rate. Adam seems to agree with Emma when pointing to the exponential function as blowing up faster.   

\subsection{Episode: Conceptual physics \label{Ex:3}}

The Instructor asks the students to solve the Schr{\"o}dinger equation to find the wave function for the free particle. Robert begins to write down the time independent Schr{\"o}dinger equation ($\hat{H}\psi=E\psi$). At this point Alex states that the equation written by Robert is time independent. On the other hand, Robert seems certain that the wave function is time independent. Robert pauses writing, and both of the students start a conceptual discussion before continuing any algorithmic manipulations. Their arguments are based upon reasoning and discussion rather than working out specific equations:
\begin{description}
\item[Robert] I think\dots.
\item[Alex] That's time independent\dots
\item[Robert] Yeah\dots Why do we need time? hmm?
\item[Alex] Hmm\dots Because the wave function might have it.
\item[Robert] If there is no force then\dots um\dots why would anything about the wave function change over time?
\item[Alex] Because the wave function might depend on time\dots from its initial condition.
\item[Robert] I don't think it did. At least\dots [unintelligible].
\item[Alex] Oh, okay.
\end{description}

Robert refers to the properties of the physical system to reason that the wave function is time independent because there is no force acting on it, whereas Alex has doubts about how the initial condition can affect the evolution of the system over time. However, Alex does not have enough evidence to justify his reasoning. 

In this episode, both students are in a conceptual physics frame, justifying their reasoning (albeit briefly) with arguments about physical quantities instead of mathematics or procedures, and bringing in more expansive reasoning. Alex thinks in terms of the feature of the problem by mentioning the initial condition. Thinking about the features of this problem also helps Robert to set the force equal to zero. This helps him to think more deeply about the underlying concepts and justify a zero change of the wave function over time.  
  
\subsection{Episode: Algorithmic physics}

In contrast to the prior example, here Robert and Alex shift into a more algorithmic frame. 
Robert continues to rearrange the equation of the Hamiltonian into kinetic and potential energy. He then sets the value of the potential energy equal to zero, and continue to recall the physical equations for the momentum based on the velocity and substitute them into the equations.

\begin{description}
\item[Robert] (writing math as he speaks) Whole definition is\dots $T+U $\dots Zero (crosses out U) is, uh $1/2$\dots  $mv$ squared. This is $p$\dots $\frac{1}{2}pv$\dots and $v$ equals $\dot x$
So H is $1/2p\dot x$
\item[Alex] You can have that $H$ equals\dots or $T$ equals $\frac{p^2}{2m}$. It's skipping all of this.
\end{description}

Robert goes through multiple steps of algebra to remember the other physics equations in order to relate the kinetic energy to the momentum. However, Alex directly recalls the equation of the kinetic energy in terms of the momentum.  Alex's framing is distinguished from algorithmic mathematics because he's not performing mathematical manipulations, merely recalling general physics formulae.  Robert is also in a recall mode, as evidenced by the words which start his observation: ``whole definition is''.  Both of them together implicitly agree that the goal of this part of the interaction is to lay out physical laws using mathematical formalism, not to discuss the applicability of those laws or derive them from first principles, as evidenced by Alex's comment that they can ``[skip] all of'' Robert's more elaborate efforts.

\section{Accounting for frame transitions}

The idea of the math-physics-algorithmic-conceptual framework is itself a development in how we model student thinking about math in physics contexts.  However, only the briefest of problems (Heller \& Heller's ``exercises''\cite{Heller1992a}) require only one frame to solve them.  To better model longer problems in upper-division physics, we must look at how students transition between frames in the course of problem solving. 
Frame transitions -- or inability to transition -- in students' problem solving illuminates the connections among ideas and procedures in longer problems.  

We identify transitions by first identifying preceding and following frames. The transition, definitionally, occurs between two different frames. Broadly speaking, we notice that the timing of transitions is relatively short (on the order of a few seconds, less than length of a few turns at talk). 


\subsection{Example: Conceptual math to algorithmic math\label{sec:Eric}}
In the following example, the group transitions from a conceptual math frame to an algorithmic frame, which is a move from an expansive to a narrow frame.

In the previous class session, students discussed that the probability density of a stationary state is time independent. Immediately prior to this example, the instructor asks the class to work in groups and find if the probability density of a superposition of two stationary states ($\Psi_{1}$ and $\Psi_{2}$) is time dependent or independent.

Eric decides to talk through the solution to the problem with his group. His preference is to start in the conceptual math frame by comparing this problem to previous problems, and his initial conclusion is that the probability density is time independent.

\begin{description}
\item[Eric]  I think\dots Cause when you do the, um, absolute value, you have to multiply by the complex conjugate, so I'm pretty sure that e thing [complex exponential part of the wave function] will just go to one, cause you'll replace that with\dots that e to the minus blah blah blah with e to the plus blah blah blah, and then when you multiply the\dots 1 over, you know, $x$ over $x$.  That's what I'm thinking.
\end{description}

We believe Eric is in the conceptual math frame because he uses reasoning based on the behavior of the exponential function and complex conjugate to determine the ``form'' of the answer; namely, that the complex conjugate causes complex exponential terms to drop out when multiplied together.  He isn't working on an algorithmic solution; he's arguing from the nature of these functions that his solution is reasonable.  After he outlines the reasoning behind his conclusion, he begins working this problem out to check his answer.

After about two minutes the instructor mentions that the answer is time dependent, which confuses Eric, who proclaims his violated expectation loudly. 

\begin{description}
\item[Eric] It is time dependent? Why? (While the instructor is explaining, he works on his paper) There's cross terms! Stupid\dots (smacks himself on forehead) ugh\dots That's why. Okay. Ugh, so stupid.
\end{description}

(From Eric's tone of voice, we interpret that Eric uses ``stupid'' to mean that his reasoning was thoughtless, not that he is personally stupid.)

The instructor's answer violates Eric's previous conclusion, prompting him to shift to another frame to explain the new answer. He realizes that his conceptual shortcut that exponentials will cancel with each other caused him to make a mistake. By viewing the problem algorithmically, Eric is able to review his work and determine what went wrong. He tries to find an answer for his question by transitioning to the algorithmic math frame and noticing that the ``cross terms'' are non-zero in this case.

\subsection{Example: Algorithmic physics to algorithmic math}

The next example illustrates a transition from algorithmic physics to algorithmic math. Both of these frames are narrow and actions in them are taken rapidly and frequently.

In this session the instructor asks the students to find the wave function of the free particle by solving the Schr{\"o}dinger equation.  
Emma begins to write down time dependent Schr{\"o}dinger equation (TDSE). She tries to remember where to put $\hbar$ in the time dependent side of the equation and asks Eric if he remembers. Eric then recalls and writes the TDSE on the whiteboard ($\frac{-{\hbar}^2}{2m}\frac{\partial^2\Psi}{\partial x^2} =i\hbar\frac{\partial\Psi}{\partial t}$). After they both finish writing the equations, Emma compares them.

\begin{description}
\item[Emma] This is what I have, good we agree. (very quickly reviewing the facts) F= 0, V=0, \dots Separable.
\end{description}

Emma is in an algorithmic physics frame, recalling equations and matching them term-by-term in preparation for solving the problem using a known procedure: separation of variables. 

However, Eric is in a different frame.  He bids to begin their problem solving by thinking about the physical system:

\begin{description}
\item[Eric] So this is like the infinite [square well], except for we don't have boundaries.
\end{description}
 
Eric's comment compares the current problem to a well-understood system and provides opportunities for further thinking about their current system. Here Eric is using reasoning about a physical system by analogy to a previous problem, which is indicative of conceptual physics thinking. However, Emma does not take up Eric's bid to use the conceptual physics frame.

After Eric's comment, Emma asks Adam to join their group.  Adam's involvement transitions the group into the algorithmic math frame, picking up Emma's earlier work on algorithmic physics to state known equations.

Adam begins by checking that the conditions for this problem satisfy the time independent Schr{\"o}dinger equation. He spends a short time in algorithmic math frame to justify that the energy in the spatial part of the Schr{\"o}dinger equation could be anything. This makes him ready to solve the bulk of the problem algorithmically and in a math frame via separation of variables.

Adam leads the group through his solution, which takes about five minutes. Adam has already taken a course on partial differential equations from the math department, and he feels very comfortable with this mathematical procedure. Adam's process begins in algorithmic math with finding the general form for $\Psi$ and the separation constant $K$. Adam uses the letter $\lambda$ as a known constant in the solution and explains it's relation to the separation constant $K$.

\begin{align}
\ \hat{H}\psi=E\psi \label{eqn:4}\                                 
\end{align}

\begin{align}
\frac{-{\hbar}^2}{2m}\frac{\partial^2}{\partial x^2} \Psi = E\psi\label{eqn:5}\
\end{align}

\begin{align}
\ {k^2}=\frac{-2mE}{{\hbar}^2}
\end{align} 
 
 \begin{align}
\ \psi''=\frac{-2mE}{{\hbar}^2}\psi
\end{align} 

\begin{align}
\ {k}=\frac{i}{\hbar}\sqrt{2mE}
\end{align} 
 
\begin{align}
\ \psi=Ae^{{\lambda}x} + Be^{-{\lambda}x}
\end{align}  
 
\begin{align}
\ \psi=Ae^{\frac{i}{\hbar}\sqrt{2mE}} + Be^{-{\frac{i}{\hbar}\sqrt{2mE}}}
\end{align}   
 
 \begin{description}
\item[Adam and Emma] $K$ squared equals minus E\dots
%
%
%
\item[Emma] (interjecting) and then we do the e to the $\lambda$ sign and then we find $\lambda$%
%
%
\item [Adam and Emma] K squared equals minus E\dots
\item[Emma] (interjecting) and then we do the e to the  sign
and then we nd \dots So now we want to find out what $\lambda$ is
\item [Adam] It is just square root of $K$ (Adam points to equation 8). No wait it's $K$, if we define the square [of] $K$ by that\dots (Adam points to equation 6)
\item [Emma] It is k\dots yea
\item [Adam] So our space part is just $e$ to the $i$, square root of $2mE$ over $\hbar$\dots (Adam writes equation 10)
\item [Emma] Just write $K$\dots Just write $K$\dots 
\item [Adam] Plus $B$ to the\dots
\item [Emma] Why would you not just write as $K$.
\item [Eric] Because this is the real name\dots I don't [unintelligible] (Adam finishes writing equation 10)
\end{description}

The students appear to stay in algorithmic math after Adam transitions them there: there is no discussion about how functions behave, what the physical system looks like, the effects on the final solution, or whether this makes sense. The students are purely focused on how to define $K$, the separation constant, and its relationship to $\lambda$, a known constant in the solution. This episode shows that algorithmic math can be a quick and powerful problem-solving mindset, but without other quadrants it is quite difficult to check if solutions makes sense.

\subsection{Example: Algorithmic math to algorithmic physics\label{sec:Adam}}

The group in the previous example continued their calculations to find out the solution of the wave function for the space part and time part.  They find that the space part is equal to $Ae^{kx}$ + $Be^{-kx}$.
and the solution of the wave function for the time part as $e^{-i(\mu/\hbar){t}}$. The next step in their algorithmic calculations is to find how $\mu$ is related to $k$. They multiply both functions and substitute the ``whole thing'' into the TDSE.  Earlier in their solution, they have found $k^2$ as $\frac{-2mE}{{\hbar}^2}$.
Part of this calculation happens while they are writing on the part of the board that is not visible in the camera. 
As they are taking the derivatives with respect to time and space, they forget a sign, which they will not notice it until later in their problem solving session (Section \ref{sec:AM}).

\begin{description} 
\item[Adam] So just $\mu$ equals ${\hbar}^2$ over $2m$ times $k^2$ [$\mu=(\frac{{\hbar}^2}{2m}){k^2}$].
\item[Eric] What's $\mu$?
\item[Emma] $\mu$ was our constant from when we were doing this part (pointing to the time derivative part of the TDSE).
\item[Eric] For time?
\item[Emma and Adam] Yea.
\item[Emma] Because we have $e^{-i(\mu/\hbar){t}}$.
\end{description}

While Emma is explaining to Eric where the coefficient $\mu$ comes from, Adam plugs in the value of $k^2$.  However, he makes an error in the denominator, and only writes $\hbar$ instead of $\hbar^2$. Emma and Adam continue with algorithmic simplifications. This mistake causes their final $\mu$ to have an extra coefficient of $\hbar$.

\begin{description} 
\item[Adam] So, boom\dots boom.
\item[Emma] cancel \dots cancel\dots cancel  \dots cancel. We get \dots {$\hbar$}{E}
\item[Adam] Hah \dots (writes ${-E}{\hbar}$ and taps his finger on the board)
\item[Emma] Is that wrong? 
\item[Adam] \dots Yeah.
\item[Emma] Because you have the $\hbar$ \dots
\item[Adam] \dots Joules (pointing to the $E$) \dots joules-second (pointing to the $\hbar$).  Yeah I don't know if that's right.
\end{description}

 Emma and Eric start algorithmic checking on the other side of the board, while Adam is silent after checking units by doing dimensional analysis.
 At this point, we can hardly hear the conversation between Emma and Eric, since the instructor has paused the problem solving session and is giving feedback to the class. The group ignores the instructor's explanation and continue to work quietly on their own.
 
\begin{description} 
\item[Emma] Oh wait  \dots wait. Isn't that \dots ?
\item[Emma] No $\hbar$ is here (pointing to the $e^{-i(\mu/\hbar){t}}$ to explain something to Eric).
\item[Emma] The $\hbar$ is here (pointing to the $\hbar$ in the time solution), and that will cancel with this one (pointing to the extra $\hbar$ in the final answer)
\item[Emma] And then we will have \dots
\item[Adam] Oh wait a minute \dots hold on \dots hold on \dots
\item[Adam] Wait \dots I know what I did \dots I did not square this (pointing to the $\hbar$ in the denominator).
 \end{description}

Adam adds the missing $\hbar$ power to his solution, and cancels it with the remaining $\hbar$ in the final answer.

\begin{description} 
\item[Adam] So it's just minus E \dots Yes \dots (while raising his fist in a triumphant gesture).
 \end{description}

In the previous episode we showed Adam's participation caused the whole group to shift from algorithmic physics to algorithmic math. In this episode only Adam shifts.  He transitions from algorithmic math to algorithmic physics to find the source of his error by thinking in terms of units of the physical quantities. In contrast to Emma and Eric, Adam does not recheck his derivatives. Instead he checks units to make sense of his answer as a physics quantity and not just a symbolic answer.  After finding the source of his error he continues the simplifications in the algorithmic math frame to find the final answer for $\mu$. This episode shows that by coordinating multiple frames students can better monitor their calculation process, saving time and/or making sense of their final answers.   

%
 
\section{Discussion}

In this study we identified the state of students' thinking associated with four discrete frames including algorithmic math, conceptual math, conceptual physics and algorithmic physics. We presented several examples of students' group problems solving switching frames to productively and correctly solve a problem. 

While upper-division students are generally facile at problem solving, on occasion they get stuck.  By observing students' behaviors we noticed moments that students change the nature of their activities to make a decision that affects the future of their problem solving, to find the source of an error in their solution, or to get ``un-stuck''. 

Epistemological framing is a window to individual's implicit state of thinking. This internal state can alter as a result of interaction with external artifacts in the environment such as the instructor's framing\cite{Irving2013framing, Nguyen2016StuFraming}.  ``Eric's''(Section \ref{sec:Eric}) shift from conceptual math to algorithmic math is responsive to the instructor's correct answer to the class. In group problem solving, shifts can also be internal to the group: when members of a group disagree, one student might cause the group to shift to another frame\cite{Nguyen2016StuFraming}. Even in individual problem solving, students may shift to another frame in the ordinary course of solving a problem. 


Epistemic games have been previously used for studying problem solving behaviors at the introductory level\cite{Tuminaro2004a}. However, at the upper division the strict move structure of these introductory e-games breaks down, and it may be more productive to look at which frames students operate in\cite{Bing2009,Irving2013framing,Nguyen2016StuFraming}.  

In a similar contrast, 
Sherin\cite{Sherin2001} compares the conceptual schemata associated with symbolic forms with Larkin's\cite{Larkin1980} ``principled-based schemata view''. He explains that the goal of his schemata is conceptual understanding and the goal of the latter schemata is step-based problem solving. However, these two views are again situated at the introductory level where conceptual mathematics is rare.  In our framework, both conceptual understanding and algorithmic thinking can be mathematical or physical, allowing for greater freedom in modeling upper-division student thinking.  As evidenced by ``Eric'' (Section \ref{sec:Eric}), conceptual thinking in not the only productive aspect of thinking about physics. 

``Eric" (Section \ref{sec:Eric}) switched from conceptual math to algorithmic math, we do not mean to imply that algorithmic math is universally more productive than conceptual math. Rather, what counts as productive framing depends strongly on the problem context, and different frames may be productive at different times within a problem.  Students' difficulties in quantum mechanics -- such as thinking that the probability density is time independent for a superposition of two stationary states, as this student does -- may simply be the result of unproductive framing and not fundamental inability to solve these problems or conceptual ``difficulties''\cite{Singh2015, Emigh2015, Passante2015}.  Modeling students' problem solving as movement in the math-physics-algorithmic-conceptual plane allows for a richer description of students' problem solving behavior than mere difficulty identification, even as difficulty identification may more exactly specify the particular confusion or incorrect reasoning students exhibit. 

There is another external factor that is more important in influencing students' framing in a problem solving context, even before being affected by other humans such as groupmates or the instructor. As soon as students read a problem, the problem statement framing interacts with the students' framing. In future work, we will use this theoretical framework to categorize students' framing and then analyze their responses as an artifact of the problem statement and not just due to the final correct answer or correct reasoning path.

\section{Conclusion}

In this study our goal was to examine students' problem solving behaviors in the often-messy setting of the classroom. We're particularly interested in how students solve problems collaboratively in groups, and especially in the ways they connect math and physics reasoning to solve upper-division problems.  
 
We identified four epistemological frames: algorithmic math, conceptual math, conceptual physics and algorithmic physics. We presented several examples of students' group problems solving switching frames to productively and correctly solve a problem. This framework divides possible student errors into three different categories as displacement error, transition error and content error. The displacement error reveals students unproductive frame of the situation. Content error shows what pieces of knowledge have to be activated to understand all the ideas incorporated in the problem frame. The last error is transition, where students have ideas in different mental spaces but do not coordinate them. 

This framework developed as a result of analysis of spontaneous and natural moments of in-class activities during one semester. There was no constraint in the problem solving sessions of the class, except, that the time duration of the problem solving session was limited. However, still students had enough time to illustrate natural moments of problem solving, to become creative, to get enough engaged with the problem to ``get stuck'' and then ``un-stuck'', or become so deeply engaged in the group problem solving to ignore the instructor's comment for several minutes after he has already paused the problem solving session.  This kind of problem solving is more ecologically valid than the problem solving in individual clinical interviews\cite{Russ2012}, and thus as a field we should attend more carefully to it. 

Instructionally, this framework is a useful tool for instructors to assess and facilitate different moments of problem solving sessions in their classroom settings. Some students like ``Eric''(Section \ref{sec:Eric}), are self reflective and get ``un-stuck'' by noticing the missing parts of their solution. However, not all of the students might act as ``Eric'' does.  Using this framework may help instructors notice when students are stuck because of unproductive framing, and give them tools to nudge students into a more productive frame.  Instructors can tip students into different frames\cite{Irving2013framing,ChariInsFraming} or gently nudge students to use additional resources\cite{Singh2015,Vygotsky1978} to resolve content errors.

\begin{acknowledgments}

The authors gratefully acknowledge the contributions of the KSUPER group who participated in inter-rater reliability testing and codebook development discussions.  An earlier version of this theoretical framework appeared in the ICLS proceedings\cite{Thompson2016ICLS}, and we are grateful to those anonymous reviewers for their feedback on the theoretical framework.  Portions of this research were funded by NSF DUE-1430967, the KSU Office of Undergraduate Research and Creative Inquiry, and the KSU Physics Department.  

\end{acknowledgments}

%

\end{document}